\documentclass[12pt]{revtex4-1}
\usepackage{amsmath,amssymb}
\usepackage[]{graphicx}
\begin{document}
%%    The information for the title page will be placed between
%%    \begin{document} and \maketitle. The order of most entries
%%    is determined by the class file and can not be changed by
%%    rearranging them. The maketitle command follows after the
%%    abstract.
%%
%%    Most of the following commands will be completed by the publisher.
%%
%%    The copyrightyear is defined in the .clo file as the first argument
%%    of the copyrightinfo command. If the copyrightyear differs from that
%%    value it might be adjusted by the following definition:
%%
%% \renewcommand{\copyrightyear}{2003}% uncomment to change the copyrightyear.
%%
%\DOIsuffix{theDOIsuffix}
%%
%% issueinfo for header and copyright line
%\Volume{55}
%\Issue{1}
%\Month{01}
%\Year{2007}
%%
%%    First and last pagenumber of the article. If the option
%%    'autolastpage' is set (default) the second argument may be left empty.
%\pagespan{3}{}
%%
%%    Dates will be filled in by the publisher. The 'reviseddate' and
%%    'dateposted' (Published online) entry may be left empty.
%\Receiveddate{15 November 2007}
%\Reviseddate{30 November 2007}
%\Accepteddate{2 December 2007}
%\Dateposted{3 December 2007}
%%
\keywords{Gauge/gravity duality, emergent geometry, black holes}

%% \pretitle{Editor's Choice}

%% We have a short and a long form for the title. The short form
%% (optional argument) goes into the running head.

\title[Sketches of emergent geometry]{Sketches of emergent geometry in the gauge/gravity duality}

%% Please do not enter footnotes or \inst{}-notes into the optional
%% argument of the author command. The optional argument will go into
%% the header.  If there is only one address the marker \inst{x} may be
%% omitted.

%% Information for the first author.
\author{David Berenstein} %

\address{Physics Department, University of California, Santa Barbara, CA 93106 }

\newcommand\vev[1]{\langle #1\rangle}
\newcommand\ket[1]{| #1\rangle}
\newcommand\bra[1]{\langle #1|}

\begin{abstract}
In this paper three notions of emergent geometry arising from the study of gauge/gravity duals are discussed. The unifying theme behind these notions of emergent geometry is that one can derive properties of the effective action of a probe or excitation around some configuration in gauge theory which can be argued to be localized at a particular position in the gravity dual, and match this description to various degrees of accuracy in the gravity dual.  The three examples discussed are giant gravitons in $AdS_5\times S^5$, open strings stretching between these giants and the probe dynamics of a D0 brane in the presence of a thermal matrix configuration of the BFSS matrix model.
\end{abstract}
%% maketitle must follow the abstract.
\maketitle                   % Produces the title.

%% If there is not enough space inside the running head
%% for all authors including the title you may provide
%% the leftmark in one of the following three forms:

%% \renewcommand{\leftmark}
%% {F. Author: A short title}

%% \renewcommand{\leftmark}
%% {F. Author and S. Author: A short title}

%% \renewcommand{\leftmark}
%% {F. Author et al.: A short title}

%% \tableofcontents  % Produces the table of contents.
\section{Introduction}

Quantum theories of gravity are non-renormalizable in dimensions higher than two. Because Lorentz invariance permits us to do arbitrarily high energy collisions by boosting the initial states, we need a UV completion of the theory of gravity.
It is possible that such a theory might be written in terms of completely different variables. We will call such a complete theory a microscopic theory of gravity. 
 In such a setup, it is not even clear that one can understand the theory geometrically 
 any more: geometry -- in the sense of having a space-time that solves Einstein's equations -- might only be an effective description around certain semi-classical configurations in the full 
 quantum theory. If we imagine such a theory that is not geometric {\em a priori}, then geometry must be a derived concept and it must emerge from the detailed dynamics of the quantum system that one 
 is trying to explore. In such a setup, all the geometric properties that one takes for granted need to be derived. For example, the existence of a metric needs to arise from the dynamics. 
 But it is not just the metric, the very idea of a causal structure or the equivalence principle need to be derived from first principles. Once one retreats enough from these familiar structures, one realizes
  that deriving geometry can be a very tall order. In such a setup, we would claim that since geometry and spacetime are derived from some more fundamental concepts (the UV complete microscopic theory), it is an emergent property of the dynamics of the theory. Indeed, considering Hawking's information loss argument \cite{Hawking:1974sw}, one can argue that even quantum mechanics is suspect.

The discovery of the gauge/gravity duality \cite{Maldacena:1997re} has been a profound revolution in our thinking both about gravity and about 
quantum field theory. This is one of the most interesting developments in that it seems to provide the first set of examples of UV complete theories of quantum gravity in dimensions larger than two. In these setups, quantum mechanics is 
preserved and the black hole information loss problem is solved in principle. The semiclassical theory of gravity and the quantum field theory which is dual to it live in different dimensions. The gravity description 
has more dimensions. These extra dimensions are not apparent in the field theory description which is dual to it. Indeed, this is one of the examples 
where the UV theory as described above can be considered not geometric and the
geometry needs to be derived. How to derive the geometry is not understood in general. There are many examples where dual pairs of field theory/ gravity solution are known, but that does not mean that we know 
how to solve the field theory and obtain the geometry.  

The purpose of this article is to give a description of recent attempts that have been made by the author and collaborators to try to understand how one should approach this problem. The main reason to 
call them a sketch is that they refer mostly to a program that suggests a way forward, but such a program is still in its infancy.

The main idea that will be explored in this paper is that to define a geometric locus, a notion that defines a `here and now', we need to be able to make an excitation, or add a probe,  and put it in place `here and now' and  at the same time we need to be able to study the dynamics of such an excitation. This is  straightforward in gravity because the theory is geometric. All one needs is a coordinate description of the geometry to be able to compute. The goal here is to do that same exercise in the field theory dual.
 The main sense in which we will study the dynamics is by trying to find an effective action  that describes the properties of the excitation/probe in terms of the variables that define the `here and now'. The variables that describe these notions are collective coordinates of the collection of quantum states that one has identified as representing the excitation. These can be considered as a coherent state representation of the quantum state of the system. The purpose is then to find an effective action for such collective coordinates.
In principle, if one finds an effective action one can then proceed and compare it to a semiclassical or classical computation in the gravity theory.  This would be the gravity prediction and at the same time one could encode how that prediction gets worse when the gravity theory becomes strongly curved by having the field theory calculation at hand. Because this is rather hard to do in general, the approach I will follow will first study a very supersymmetric set of states: giant gravitons in $AdS_5\times S^5$.
These preserve half of the supersymmetries, and because of the large amount of remnant supersymmetry they are much more protected from receiving corrections than other states. In particular, all such half BPS states can be classified completely both at weak coupling in field theory \cite{Corley:2001zk} in terms of Young tableaux and in the gravity theory \cite{Lin:2004nb} in terms of bubbling geometries. The two classifications can be related to each other using a free fermion representation
of the states \cite{Berenstein:2004kk} which shows that they can be understood in terms of fermion droplets for the integer quantum hall effect. The  point to be discussed is how to extract precise geometric features of the giant graviton dynamics from the field theory states directly. 
Then I will proceed to study open strings stretching between such giants and show how different notions of geometry for strings and branes are actually compatible with each other. More to the point, I will show 
that this compatibility ends up being related to Local Lorentz invariance in the gravity dual.
At the end of the paper I will describe some progress towards understanding geometric aspects of matrix black holes (thermal matrix configurations) in the BFSS matrix model \cite{Banks:1996vh} and the ideas this suggests for the interior black hole dynamics, as well as how the notion of a horizon can appear dynamically in such a setup.

\section{A collective coordinate and a giant graviton effective action}

The first example I will discuss is the problem of finding an effective action for a giant graviton in $AdS_5\times S^5$. I will study this problem by using the field theory dynamics of ${\cal N}= 4 $ SYM, rather than the gravitational dual. A giant graviton on $AdS_5\times S^5$  \cite{McGreevy:2000cw} is a half-BPS state that is represented by a D3-brane wrapping an $S^3\subset S^5$ and which is at the origin in $AdS_5$. This is a statement about where the brane is located in global coordinates for $AdS_5$, which can be represented as $ds^2 = d\rho^2 -\cosh^2\rho dt^2 + \rho^2 d\Omega_3^2$. The origin is the set $\rho=0$, whereas the $d\Omega_3^2$ is a round sphere metric for an $S^3$ slicing of $AdS_5$. 
 The state preserves an $SO(4)\times SO(4)$ symmetry. One of these is the isometry of the $S^3$ that the brane is wrapping. The other one is the $SO(4)$ rotational symmetry of the $S^3$ sphere in the 
$AdS_5$  geometry above. 
The brane moves on the $S^5$ so that it's net angular momentum is equal to the energy of the state. The field theory duals of such giant graviton states were proposed in \cite{Balasubramanian:2001nh}, and more generally, the half BPS states in the ${\cal N}=4$ SYM field theory can be classified exactly \cite{Corley:2001zk}. These generally admit a description in terms of fermion droplets for a gas of charged particles in a magnetic field in two dimensions \cite{Berenstein:2004kk}. We will adopt this language of the droplet in what follows.

For convenience, the $S^5$ can be pictured as an $S^3$ fibration over a disc. To do this we consider the split in six dimensions into $2+4$, and writing the set of 4 coordinates that has been split into spherical coordinates. 
\begin{eqnarray}
ds^2 &=& (dx^1)^2 + (dx^2)^2+ \sum_{i=1}^4(dy^i)^2 \\
&=& (dx^1)^2 + (dx^2)^2+ dr^2+r^2 d\Omega_3^2
\end{eqnarray}
The $S^5$ sphere is obtained by restricting to $r^2=1-x_1^2-x_2^2$ the above. A giant graviton wrapping the $S^3$ will be described by the two coordinates $x^1, x^2$, where we will have constant rotational motion in the $x_1,x_2$ disc. This is, the coordinates will move at constant angular speed $x_1+ix_2 = s \exp(i t+i\theta)$ and stay at fixed radius \cite{McGreevy:2000cw}. What one realizes is that if we fix the classical energy and angular momentum (we fix the value of $s$), we still have a one parameter family of such solutions characterized by the phase $\theta$. This is to be considered as a zero mode of rotations, which is dual to angular momentum. If we quantize the system at fixed angular momentum, we should get a wave function for $\theta$ describing a state that is smeared at all values of $\theta$. This is not what a classical state should look like. A classical state should be localized
in both coordinates. This becomes clear if we have two such objects and we want to stretch strings between them (after all, giant gravitons are D-branes): the energies of those strings depend on the angular separation between two such branes, not just the difference in angular momentum (the two values of $s$). Thus if we try to solve for the spectrum of these strings in the field theory dual, we need to resolve the angular separation between the giant gravitons by resumming a problem in degenerate perturbation theory. This is what was done in a series of works culminating in \cite{Koch:2011hb}. Alternatively, we can find a description where this geometry of the localized angle is resolved from the beginning by looking at the correct states that are localized both in angle and radius. My goal is to explain how to undertake this second path by finding a collective coordinate description of the giant gravitons in the dual field theory, following \cite{Berenstein:2013md}. 

The basic idea is as follows. The boundary of $AdS_5$ in global coordinates is a cylinder $S^3\times R$. Any state in the bulk can be represented by a quantum field configuration on the $S^3$ spatial boundary.
For the ${\cal  N}=4$ conformal field theory at weak coupling the description of the quantum states in the $S^3$ can be done in terms of a Fock space of free fields supplemented by the Gauss' law constraint.  And then, via the operator-state correspondence, one can in turn represent such states as an insertion of a local operator at the origin. The operators in question are made only of the s-wave of the complex field  $Z$, so they can be described as local polynomials of multi-traces of $Z$ without any derivatives. The $Z$ field is
one of the complex scalar fields  that make the lowest component of a chiral field of ${\cal N}=4 $ SYM when decomposed in terms of an ${\cal N}=1$ superspace formulation.

In \cite{Balasubramanian:2001nh} the giant graviton states are built out of subdeterminant operators $\det_\ell(Z)$, but they have fixed angular momentum $\ell$. The norm of these states is
\begin{equation}
\langle \det\!_\ell \bar Z \det\!_\ell Z\rangle= \frac{N!}{(N-\ell)!}
\end{equation}
and this is the normalization according to the Zamolodchikov metric. The ansatz for the collective coordinate description of a giant graviton is the following
\begin{equation}
\det(Z-\lambda)= \sum_{\ell=0}^N (-\lambda)^{N-\ell}\det\!_\ell(Z)
\end{equation}
where the complex parameter $\lambda$ defines a very special family of states in the field theory. 
This is equivalent to the following quantum mechanical superposition of states
\begin{equation}
|\lambda\rangle= \sum_\ell (-\lambda)^{N-\ell} |\ell\rangle 
\end{equation}
where we identify $|\ell\rangle\simeq \det_\ell(Z)$ under the operator state correspondence, including the normalization of the states.
A straightforward computation shows that 
\begin{equation}
\langle \det(\bar Z-\tilde \lambda^*) \det(Z-\lambda)\rangle = N!\sum_{\ell=0}^N \frac{(\lambda\tilde \lambda^*)^\ell }{\ell!} \simeq N! \exp(\lambda\tilde\lambda^*)= \langle \tilde \lambda | \lambda\rangle
\end{equation}
where the approximation to the exponential is valid so long as $\lambda\tilde \lambda^*< N-O(\sqrt N)$. The parameters $\lambda$ have a size that is bounded by $\sqrt N$ and live on a disk. 
We can use these states to compute an effective action for the collective coordinate $\lambda$. We do this by approximating solutions of the Schr\"odinger equation by orthonormalized states belonging to 
our collective coordinate description, and computing a Berry phase contribution $\bra \lambda \partial_t \ket \lambda$. The end result is
\begin{equation}
S_{eff}= \int dt\left[\bra{\lambda} i \partial_t \ket{\lambda}- \bra{\lambda} H \ket{\lambda} \right] = \int dt \left[\frac i 2   (\lambda^* \dot \lambda - \dot \lambda^* \lambda) - ( N-\lambda \lambda^*)\right]\label{eq:effaction}
\end{equation}
Solving this effective action gives an inverted  harmonic oscillator in a first order formalism, and results in $i\dot \lambda= - \lambda$, or $\lambda= \lambda_0 \exp(it)$. Also, notice that if $|\lambda-\tilde \lambda |>1 $, the overlaps for normalized states are exponentially suppressed. This means that the coordinates $\lambda$ give rise to a reasonable geometric space, with a bit of quantum uncertainty. This uncertainty is of order one, while the droplet is of size $\sqrt N$, so in the large $N$ limit we get arbitrary localized states relative to the size of the droplet.
In this sense we reproduce the gravity result: we have a particle on a disc moving at constant angular velocity equal to one. This makes the droplet picture of \cite{Berenstein:2004kk} rather precise: it is the geometry where the giant gravitons move. It is worth pointing out that quantizing the effective action (\ref{eq:effaction}) reproduces the original Hilbert space of states. We need to be careful because $\lambda$ is restricted, $|\lambda|^2<N$. The energy of a state is $N-|\lambda|^2\geq 0$, which is in accordance with the statement that the quantum theory of fields on the $S^3\times R$ has positive energy relative to the vacuum.

Indeed, one can go further an identify the fermion droplets exactly with supergravity solutions \cite{Lin:2004nb}. This gives rise to a coordinate system on $AdS_5\times S^5$ where the incompressible quantum liquid droplets of the free field theory description get replaced by an incompressible liquid in a two dimensional submanifold of the asymptotically $AdS_5\times S^5$ geometries. The area of the liquid droplets in the gravity side count D-brane charge, and they have to be quantized. This quantization is due to Dirac's quantization conditions for the five form self-dual field of type IIB supergravity on compact five cycles. Giant gravitons are limits of such configurations. 

The result above shows that we have the topology of a disc and a Poisson bracket, but it does not show a metric property immediately. One can consider building such a metric if one assumes that the symplectic form arises from
a Kahler manifold. In that case, the metric of the $\lambda$ plane is flat.  In order to actually measure a distance, we can subtend a string between two such giant gravitons. The energy of such a string depends on the distance, and the greater the distance, the greater the energy. This is the problem we will take on in the next section.

\section{Cuntz oscillators and strings}

The process of finding operators that describe strings attached to giant gravitons was described in \cite{Berenstein:2003ah}, and for multiple giants it was formulated in \cite{Balasubramanian:2004nb}. That this procedure gives the right answer in general was proved in \cite{deMelloKoch:2012ck}. Here, I will follow the results in \cite{Berenstein:2013eya} to describe how the spectrum of strings stretching between giant gravitons 
is geometric in the collective coordinates we found above. The precise details of the computations can be found there. For convenience, we work in an orbifold of ${\cal N}=4 $ SYM. This should not affect the end result of the computation, so long as one can argue that planar equivalence between parent and daughter theories takes place consistently even in the presence of D-branes. Usually this consistency 
requires that the method of images work (as exemplified in \cite{Douglas:1996sw}), with the D-branes in question being the giant gravitons. One way to think about this is that the planar diagrams that give rise to string worldsheets and to a spin chain for the corresponding strings are the same in ${\cal N}=4$ SYM and it's orbifolds , and that the only difference is about how one closes the string by identifying the correct periodicity conditions that need to be applied \cite{Beisert:2005he}. If one studies open spin chains, there is less to check. Each boundary of the spin chain will give rise to a boundary condition that depends on the brane where the spin chain ends, and this is where understanding the images on the cover will do the trick. The important thing to note is that adding strings between giant gravitons necessarily requires using fields other than $Z$. The simplest such fields will be scalar fields, of which we pick a particular complex combination. Let us call it $Y$. The set of states built out of $Z,Y$ is called the $SU(2)$ sector. The one loop hamiltonian can be described as 
\begin{equation}
H_{1-loop} \simeq g_{YM}^2\hbox{Tr} [Y,Z][\partial_Z,\partial_Y]
\end{equation}
where factors of $2\pi$ that depend on the normalization of the fields have been ignored. This follows immediately from \cite{Minahan:2002ve} and is written explicitly in \cite{Beisert:2003tq}. We need to apply this to 
states of the form
\begin{equation}
\label{eq:openspinchainorbifold}
\det(Z-\lambda) \det(\tilde Z-\tilde \lambda)\hbox{Tr}\left( \frac{1}{Z-\lambda} Y_{12} \tilde{Z}^{n_1} Y_{21} Z^{n_2} Y_{12} \tilde{Z}^{n_3} \dots Z^{n_k} Y_{12} \frac 1{\tilde{Z} - \tilde{\lambda}} X_{21}\right)
\end{equation}
where an extra $X_{21}$ is used to return the charge from the second brane to the first one. The $Z,\tilde Z$ represent scalar field holomorphic superpartners of the gauge bosons in an $Z_2$ orbifold of $N=4$ SYM with ${\cal N}=2$ supersymmetry. When thinking of the open spin chain, we think of only the combination 
\begin{equation}
 Y {Z}^{n_1} Y Z^{n_2} Y Z^{n_3} \dots Z^{n_k} Y
\end{equation}
with the boundary conditions determined by $\lambda, \tilde \lambda$. We choose to label these states as $|\lambda, \tilde\lambda; n_1, \dots, n_k\rangle$, where we count the number of $Z$ between the $Y$, and we count each of these as a site. Therefore we have $k$ sites on the spin chain. 
This is a bosonic representation of the word above. It is convenient to define a Cuntz oscillator for each site, with $a^\dagger|n\rangle= |n+1\rangle$ and $n\geq 0$, which commute between different sites. The one loop Hamiltonian, including the contributions from the diagrams of the boundary is the following \cite{Berenstein:2005fa,Berenstein:2013eya}
\begin{eqnarray}
H_{\text{spin chain}} &\simeq &g_{YM}^2 N \left[\left(\frac{\lambda}{
\sqrt N}-a_1^\dagger\right) \left(\frac{\lambda^*}{
\sqrt N}-a_1\right)+ (a^\dagger_1-a^\dagger_2)(a_1-a_2)\right.\\&& \left.
+\dots +\left(\frac{\tilde\lambda}{
\sqrt N}-a_k^\dagger\right) \left(\frac{\tilde \lambda^*}{\sqrt N}-a_k\right)\right]
\end{eqnarray}
We see that it is a sum of squares. To solve for the ground state, it is important to consider coherent states for the Cuntz oscillators, $a \ket z= z\ket z$. These states are normalizable so long as $|z|< 1$. Thus, they live on a disk. Making a coherent state ansatz for the ground state and mimizing the energy solves for the exact ground state of the problem. The ground state is the state
\begin{equation}
\ket 0_{k, \lambda, \tilde \lambda}= \ket{\lambda, \tilde\lambda; z_1\dots z_k}
\end{equation}
with 
\begin{equation}
\frac{\lambda^*}{
\sqrt N}-z_1 = z_1 - z_2 = \dots = z_i -z_{i+1} = \dots = z_k-\frac{\tilde \lambda^*}{
\sqrt N}\label{eq:equal}
\end{equation}
Solving these equations leads to 
\begin{equation}
z_i -z_{i+1}= \frac 1{k+1}\left(\frac{\lambda^*}{
\sqrt N}-\frac{\tilde \lambda^*}{
\sqrt N}\right)
\end{equation}
where $z_0=\frac{\lambda^*}{
\sqrt N}= \xi $ and $z_{k+1}=\frac{\tilde \lambda^*}{
\sqrt N}= \tilde \xi$.
We see that if we introduce normalized coordinates for the $\lambda$ disc (represented by $\xi, \tilde \xi$ above), so that it's radius is equal to one, then these normalized coordinates enter exactly in the description of the boundary of the coherent state representation of the open string stretching between two giant gravitons. As a bonus, we get a linear system of equations for the $z_i$ which shows the string stretching uniformly between $\xi, \tilde \xi$ with the collective coordinates (coherent state coordinates $z$) having a simple geometric interpretation on the same disk as the $\xi$. 
Moreover, the one loop energy of the state is
\begin{equation}
E^{(1)}_0 = A \frac{g_{YM}^2N}{k+1}\left|\xi-\tilde \xi \right|^2 \label{eq:energyrel}
\end{equation}
where in $A$ we hide all the normalization factors we have ignored so far.
This result depends on the distance squared between $\xi$ and $\tilde \xi$. This is, we see a notion of distance in the $\xi$ plane appearing from a computation. We still need to explain the $1/(k+1)$ in front of it.
We will do this a bit later on.

 Again, it is possible to write an effective action for the spin chain above with a general coherent state ansatz at each site
 $S_{\text{eff}}=\int dt\, \bra {z(t)} i \partial_t \ket {z(t)} - \int dt\, \langle H \rangle  $
\begin{equation}
S_{\text{spin chain}} = \int dt \left[\frac{i}{2}\sum_{i=1}^k \frac{\bar{z_i} \dot{z_i} - \dot{\bar{z}}_i z_i}{(1-\bar z_i z_i)} - (k+1) - \sum_{i=1}^k \frac{ \bar z_i z_i}{1-\bar z_i z_i} 
-  A g_{YM}^2 N \sum_{i=0}^k |z_{i+1}-z_i|^2  \right] 
\end{equation}
which includes the classical dimension of the $Y$ and $Z$ fields in the second and third terms. An open string with $k+1$ $Y$ actually has $R$-charge $k+1$ in a direction orthogonal to the R-charge of $Z$. This is interpreted as angular momentum on the $S^5$ in the directions along $S^3$.
If we look at the equations of motion of the $z_i$ and $\xi,~\tilde{\xi}$ in the case $g_{YM}^2=0$, we find that
\begin{eqnarray}
\dot{\xi} = -i \xi, &\quad \dot{\tilde{\xi}} = - i \tilde{\xi} \\
\dot{z_i} =& -i z_i 
\end{eqnarray}
so all of the pieces rotate at the same uniform angular velocity on the $\xi, \tilde \xi$ disk. The one loop correction to the energy is constant for these configurations. More precisely, the total action is
\begin{eqnarray}
S_{\text{tot}} &= N \int dt \left[\frac{i}{2}(\xi\dot{\bar{\xi}} - \bar{\xi} \dot{\xi}) - ( 1-\bar \xi \xi) \right]+N \int dt \left[\frac i{2} (\tilde{\xi}\dot{\bar {\tilde{\xi}}} - \bar{\tilde{\xi}}\dot{\tilde{\xi}}) - (1 - \bar{\tilde\xi}\tilde{\xi}) \right]\\
&\qquad - \int dt \left( 1+ A g_{YM}^2 N |\xi -\tilde{\xi}|^2\right)+ S_{\text{spin chain}}
\end{eqnarray}
Because the action in terms of $\xi$ is of order $N$, we clearly see that they are $D$-branes: the prefactor is roughly the tension of the object. If we excite enough strings between the branes, the extra term $|\xi-\tilde \xi|^2$ starts competing with the free action for $\xi$ and it starts modifying the solutions to the equations of motion for $\xi$. At this order in the loop expansion, one sees that the net effect is to alter the dynamics of $\xi$ by a quadratic term. Thus one gets a modified set of harmonic oscillators. This is what was observed in \cite{Koch:2011hb,deMelloKoch:2011ci} with the result of a completely combinatorial calculation. Here we see that a spring with $k$ sites and $(k+1) $ $Y$ words is effectively a spring with constant $1/(k+1)$, so it provides for $k+1$ springs of strength one in series, whereas multiple occupation of a word defect, let us say $s$ of them produces an effect which is $s$ times larger. This is, we can build arbitrary rational spring constants by taking $s$ springs of constant $1/(k+1)$ (including $k=0$). This ignores joining and splitting of strings, which should also be taken into account. 

At higher loop orders, the quadratic effect $|\xi -\tilde \xi|^2$ above,  becomes a square root  and matches the idea of a relativistic dispersion for strings stretching between two branes \cite{Berenstein:2013eya}. 
This is, we expect that 
\begin{equation}
E_{string}\simeq \sqrt{ (k+1)^2 + 2 A g_{YM}^2 N |\xi -\tilde{\xi}|^2} 
\end{equation}
This is very important: it shows that the sphere directions of the $S^5$ and the time direction of $AdS_5$ which have different origins are compatible in the sense that one can boost objects along the $S^5$. 
When we expand this formula in the large  small $g{YM}^2N$ limit, or the large $k$ limit, we recover the result in equation (\ref{eq:energyrel}).  We see that the factor of $k+1$ in the denominator is due to a kinetic 
effect for a relativistic particle.
Notice that all of our discussion on the disk geometry is happening in the $S^5$ of the gravitational theory and the time direction is from the $AdS_5$ geometry. More precisely, the time in global $AdS_5$ is also the time direction on the boundary which is geometric on the boundary. The relativistic geometry is actually arising in a very detailed form from direct computations in the dynamics of the  ${\cal N}=4 $ SYM theory.

\section{Effective field theory probes of matrix black holes}

So far, we have dealt with a vacuum configuration in the gauge/gravity duality. Such configurations usually correspond to a horizonless solution in the gravity theory. 
Obviously, if everything that can happen in gravity can happen in the field theory dual, then the notion of a horizon must also emerge dynamically in the gravity dual. This is what will be explored next. 
The details I'm going to present grew out of numerical studies in the BFSS matrix model \cite{Asplund:2011qj,Asplund:2012tg}. The original idea was to explore the real time dynamics of a black hole in the 
dual field theory. Because simulating non-linear quantum systems in real time is very hard due to the sign problem, one could hope that a classical real time evolution might 
contain useful information to organize the problem. If one furthermore wishes to avoid the UV catastrophe without using quantum mechanics (reintroducing $\hbar$), one way to do so is to work with a finite number of degrees of freedom. One also wants such a model to be subject to the gauge/gravity duality. This is accomplished by the BFSS matrix model \cite{Banks:1996vh}. The problem that became important then was how to visualize  the random configurations of matrices that represent a state in the classical theory. This is, how to define a notion of here and now. The notion if now is straightforward: we just pick the simulation at some time $t$ and analyze such configurations. The harder problem is to define a notion of here. 

To solve that problem, we need to understand the vacuum configurations of the BFSS matrix model. The variables defining the matrix model are nine hermitian matrices $X_i$ and their superpartner fermions $\Psi$.
The potential is in the form of a commutator squared $ V(X) \simeq -\sum_{ij}\hbox{tr}([X_i,X_j])^2\geq 0$ and is positive definite. A minimal energy configuration requires that the matrices commute. Because the matrices are hermitian, if they commute they can be diagonalized simultaneously by conjugating with a unitary matrix $U$. This diagonalization by $U$ is a gauge transformation, so only the eigenvalues are gauge invariant. These are arranged into $N$ vectors $\vec x_\alpha$, which denote the eigenvalue of a joint eigenvector in each of the nine matrices. These are interpreted as the coordinates of $N$ particles in  nine flat dimensions \cite{Banks:1996vh}, 
that can be related to the transverse  directions to the lightcone in a light-like compactification of M-theory on a circle. 
When the matrices don't commute, we're still supposed to think of them as a more complicated D-brane configuration embedded geometrically in such flat nine dimensions. This suggests that to probe the location of such a brane, we need to add a probe D0-brane at some locus $\vec \lambda$ and ask what that D0-brane sees.
We are essentially  defining a proxy for the geometry: we define the geometry by reference to the probe position $\vec \lambda$. Depending on what we see  at $\lambda$, we can color the reference geometry with that information
and use that as our notion of geometry. We know that this is the correct description in the asymptotic region, which can be traced to how we understand gravitational interactions between D-brane configurations \cite{Taylor:1998tv,Taylor:1999gq}. 

To add such a probe, we need to take a matrix configuration and add an extra row and column to each matrix. The only non-zero entry that is added is the common component of the row and the column, which is set to the position of the probe. This is, we make a new configuration $\tilde X_i = X_i \oplus \lambda_i$, where $\vec \lambda$ is the position of the probe. Indeed, if $X_i(t)$ and $\lambda_i(t)$ are solutions to the classical equation of motion of the matrices and they satisfy the Gauss' law constraint (trivially for $\lambda_i(t)$), then so is $X_i(t)\oplus \lambda_i(t)$. If we do this in the classical theory, the two D-brane configurations are not seeing each other at all.
To understand what the probe D-brane sees, we need to understand the additional degrees of freedom that were added to the system when we added the probe. The idea is that to properly define the action of the $\lambda_i$, we need to include the quantum corrections induced by the additional degrees of freedom that we chose to set to zero classically above. To set these states to zero in the quantum theory we need to integrate them out and consider the question if this integrating out procedure makes sense or not. To the extent that this is a well defined  computation (at least in principle), then we are still allowed to talk about geometry. This is, to define what happens at $\lambda$, we need to introduce $\hbar$ again. We do not want to do that directly on the matrix $X$, but we can do it on the additional row and column in the {\em background} defined by $X$. Although the matrix $X$  is obtained from the classical dynamics, one can imagine that it represents a typical representative  of the matrix black hole wavefunction (the square of the wave function is a probability distribution on the $X$, from which we can pick a typical representative). 

 For our purposes, we concentrate on the fermion degrees of freedom connecting the probe to the matrix configuration. The instantaneous Hamiltonian for the fermion degrees of freedom connecting the matrix to the probe is given by \cite{Berenstein:2012ts}
 \begin{equation}
 H_{od} \simeq {\hbox{tr}}\sum_i (\bar \psi  X_i \gamma^i \psi-  \psi x_i \gamma^i \bar \psi)
 \end{equation}
where we have decomposed the off diagonal degrees of freedom according to their charge with respect to the $U(1)$ charge of the probe brane
\begin{equation}
\Psi \simeq \begin{pmatrix} 0 & \psi\\
\bar \psi &0 \end{pmatrix}
\end{equation}
Solving for the dynamics of $H_{od}$ is equivalent to solving for the spectrum of the effective Hamiltonian
\begin{equation}
H_{eff} = \sum_i  (X_i -\lambda_i)\otimes \gamma_i\label{eq:effham}
\end{equation}
This $H_{eff}$ is a hermitian matrix, so it's eigenvalues are real.
The positive eigenvalues are interpreted as raising operators for modes of $\psi$, while the negative eigenvalues are lowering operators for the $\bar \psi$ excitations. The absolute value of the eigenvalue is the energy of the mode. These modes are interpreted as strings stretching between the probe and the configuration. The lightest such string defines a notion of distance: the energy of a classical (static)  string stretched between two branes located at $A,B$ is the length of the string times the string tension. If we quantize the string, there is an additional zero point energy contribution. This additional zero point energy contribution is absent in the Ramond sector because worldsheet supersymmetry is preserved. The Ramond sector for open strings corresponds to spacetime fermions, so we find that distance is naturally measured by fermionic degrees of freedom. The notion of distance we have produced this way is based on the spectrum of \eqref{eq:effham}. For large values of $\lambda$, we can consider the $X_i$ as a perturbation, so we find that the spectrum of $H_{eff}$ has eigenvalues
$\pm|\vec \lambda|$ in equal numbers, with a first order correction which is characterized by the eigenvalues of the matrix $\vec X$ along the direction specified by $\vec \lambda$, this is,  $\hat \lambda \cdot X$ . The distance is then $\lambda - x_{\hat \lambda}$, with $x_{\hat \lambda}$ the largest eigenvalue of $\hat \lambda \cdot X$. As we come near the configuration defined by $\vec X$, the eigenvalues of $H_{eff}$ get distorted (we can think of this effect as higher order corrections in treating $X$ as a perturbation to $\vec \lambda\cdot \vec \gamma$) and we can eventually find that the eigenvalues can and do cross zero (a raising operator becomes a lowering operator). When this happens we can not integrate out the fermions. The locus where the fermions become nearly massless is the location of the configuration $X$, and this region can be considered to be gapless: there is very little energy cost on creating fermions. A rather interesting question to ask is how many fermions are nearly massless. If we imagine that a matrix black hole has a temperature $T$, then these  fermions can become active degrees of freedom if $\hbar \omega< T$ and those fermions can not be integrated out.  This gives a criterion to understand if $\omega$ is large or small relative to $T$. So long as it is small relative to $T$ we can call it a gapless mode. The region where this occurs is a ball around the origin and the gapless modes are essentially uniform over this ball.

To determine how many modes are massless, one  studies the statistics of the density of states near zero energy. This should be somewhat universal give a typical state of the configuration $X$.
It turns out that there are many such degrees of freedom. One in general expects that $\rho(\omega)\simeq \omega^{d-1}$ where $d$ is a critical exponent \cite{Berenstein:2013tya}, which we baptized as the spectral dimension of $H_{eff}$. If $H_{eff}$ was a $d$-dimensional field theory of free fermions at finite volume, the density of states would scale in the same way. The gapless region for the random configurations $X$ covers a roughly spherical ball around the origin. If this was well represented by a gas of point like D-branes with small off-diagonal excitations, then one can easily convince oneself that the natural value of $d$ is $d=9$. Measuring it directly by numerical configurations one finds that $d=1$  \cite{Berenstein:2013tya}. This is, the density of states is constant near zero and effectively over the whole set of eigenvalues of $H_{eff}$. Such a value of $d$ can not arise from local centers where strings end or a model of nearly commuting  ball filling matrices: all such models produce $d=9$. The value found in simulations is not compatible with such local physics. 

Because near black hole horizons one also gets a spectrum of strings whose energy goes to zero, one can argue that the locus where the fermions become gapless for the first time should be considered to be the horizon of the black hole. 
Away from this locus, the fermions are massive and can be integrated out. The effective field theory of the probe makes sense. Inside the gapless region we can not isolate the degrees of freedom of the position of the probe because the fermions can not be integrated out: this is especially problematic at strong coupling (large $\hbar$). At small $\hbar$ the fermions don't cause backreaction on $\lambda$ even if they get excited. But as we take $N\to \infty$, the number of such gapless modes grows with $N$, even at fixed $T,\hbar$. This will dominate the energy of the probe. One can also study the notion of a horizon in terms of the free energy or entropy of a probe in the mean field approximation \cite{Iizuka:2001cw} with qualitatively similar results: a radius where the entropy differs substantially from one indicating that strings stretching to the horizon are thermalized. The gapless region is thus excised form the geometric characterization of the outside because we are choosing to define geometry by having a reasonable effective action for a probe.
 It is interesting to speculate that this non-local dynamics is a model for the physics behind the horizon of a black hole and as such provides a physical model for {\em firewalls} \cite{Almheiri:2012rt,Braunstein:2009my} (see also \cite{Almheiri:2013hfa}).

\acknowledgments
I would like to thank my collaborators for their immeasurable contributions to the papers that make up the point of view presented here. These are C. Asplund, V. Balasubramanian, D. Correa, E. Dzienkowski, B. Feng, M.-x. Huang, D. Trancanelli, S. Vazquez. Work supported in part by DOE under grant DE-FG02-91ER40618

\end{document}